\documentclass[runningheads]{llncs}
\usepackage{latexsym}
\usepackage{graphicx}
\usepackage{amsfonts,amssymb,amsmath}
\usepackage[nolist,nohyperlinks]{acronym}
\usepackage{cite}
\usepackage{multirow}
\usepackage{color}
\usepackage{mathtools}
\usepackage[caption=false]{subfig}
\DeclarePairedDelimiter\ceil{\lceil}{\rceil}

\begin{acronym}
\acro{MIMO}{multiple-input multiple-output}
\acro{LIS}{large intelligent surface}
\acro{CSI}{channel state information}
\acro{5G}{5th generation of wireless networks}
\acro{LoS}{line-of-sight}
\acro{FC}{fully connected layer}
\acro{PDF}{probability density function}
\acro{SVM}{support vector machine}
\acro{SNR}{signal-to-noise ratio}
\acro{ML}{machine learning}
\end{acronym}

\begin{document}
\title{A Primer on Large Intelligent Surface (LIS) for Wireless Sensing in an Industrial Setting\thanks{This project has received funding from the European Union’s Horizon 2020 research and innovation programme under the Marie Sklodowska-Curie grant agreement No 813999. 
This work has been submitted to Springer for possible publication. Copyright
may be transferred without notice, after which this version may no longer be
accessible.
}  
}
\titlerunning{LIS for Wireless Sensing}
%
\author{Cristian J. Vaca-Rubio\inst{1} \and
Pablo Ramirez-Espinosa\inst{1} \and
Robin Jess Williams\inst{1} \and
 Kimmo Kansanen\inst{2} \and
 Zheng-Hua Tan\inst{1} \and 
 Elisabeth de Carvalho\inst{1} \and
 Petar Popovski\inst{1}}
\authorrunning{Cristian J. Vaca-Rubio et al.}
%
\institute{Department of Electronic Systems, Aalborg University, Denmark \\ \email{\{cjvr, pres, rjw, zt, edc, petarp\}@es.aau.dk}\and
Norwegian University of Science and Technology, Trondheim, Norway
\email{kimmo.kansanen@ntnu.no}}

\maketitle              
%
\begin{abstract}One of the beyond-5G developments that is often highlighted is the integration of wireless communication and radio sensing. This paper addresses the potential of communication-sensing integration of Large Intelligent Surfaces (LIS) in an exemplary Industry 4.0 scenario. Besides the potential for high throughput and efficient multiplexing of wireless links, an LIS can offer a high-resolution rendering of the propagation environment. This is because, in an indoor setting, it can be placed in proximity to the sensed phenomena, while the high resolution is offered by densely spaced tiny antennas deployed over a large area.  By treating an LIS as a radio image of the environment, we develop sensing techniques that leverage the usage of computer vision combined with machine learning. We test these methods for a scenario where we need to detect whether an industrial robot deviates from a predefined route. The results show that the LIS-based sensing offers high precision and has a high application potential in indoor industrial environments. 
\end{abstract}
%
%
%
\section{Introduction}
\label{Introduction}

Massive \ac{MIMO} is a fundamental technology in the \ac{5G}, with the addition of a large number of antennas per base station as its key feature \cite{andrews2014will}. 
Looking towards post-5G, researchers are defining a new generation of base stations that are equipped with an even larger number of antennas, giving raise to the concept of \ac{LIS}. Formally, an \ac{LIS} designates a large continuous electromagnetic surface able to transmit and receive radio waves \cite{Hu2018}, which can be easily integrated into the propagation environment, e.g., placed on walls. In practice, an \ac{LIS} is composed of a collection of closely spaced tiny antenna elements. Whilst the performance of \ac{LIS} in communications has received considerably attention recently 
\cite{basar2019transmission,  Dardari2019, Hu2018, Bjornson2020}, the potential of these devices could go beyond communications applications, e.g., environment sensing. Indeed, such large surfaces contain many antennas that can be used as sensors of the environment based on the \ac{CSI}.

Sensing strategies based on electromagnetic signals have been thoroughly addressed in the literature in different ways, and applied to a wide range of applications. For instance, in \cite{wang2016rt}, a real-time fall detection system is proposed through the analysis of the communication signals produced by active users, whilst the authors in \cite{pu2013whole} use Doppler shifts for gesture recognition. Radar-like sensing solutions are also available for user tracking \cite{zhao2013radio} and real-time breath monitoring \cite{adib2014real}, as well as sensing methods based on radio tomographic images \cite{zhao2018through, wilson2010radio}. Interestingly, whilst some of these techniques resort solely on the amplitude (equivalently, power) of the receive signals \cite{zhao2013radio, wilson2010radio}, in those cases where sensing small scale variations is needed, the full \ac{CSI} (i.e., amplitude and phase of the impinging signals) is required \cite{zhao2018through, adib2014real}.  

On a related note, \ac{ML} based approaches are gaining popularity in the context of massive \ac{MIMO} systems,  providing suitable solutions to optimization problems \cite{joung2016machine, Demir2020, Ma2020, Huang2018}. Due to the even larger dimensions of the system in extra-large arrays, deep learning may play a key role in exploiting complex patterns of information dependency between the transmitted signals.

The popularization of \ac{LIS} as a natural next step from massive \ac{MIMO} gives rise to larger arrays and more degrees of freedom, providing huge amounts of data which can feed \ac{ML} algorithms. Hence, deep learning arises as a potential solution to exploit the performance of \ac{LIS}.  

In this work, we aim to pave the way to the combined use of both deep learning algorithms and the aforementioned large surfaces, exploring, for first time in the literature, the potential of such a joint solution to sense the propagation environment. Specifically, the contribution of this work is twofold:
\begin{itemize}
    \item We propose an image-based sensing technique based on the received signal power at each antenna element of an \ac{LIS}. These power samples are processed to generate a high resolution image of the propagation environment that can be used to feed computer vision algorithms to sense large-scale events. 
    \item A computer vision algorithm, based on transfer learning and \ac{SVM}, is defined to process the radio images generated by the \ac{LIS} in order to detect anomalies over a predefined robot route.
\end{itemize}

The performance of the proposed solution is tested in an indoor industrial scenario, where the impact of the array aperture, sampling period and the inter-antenna distance is thoroughly evaluated. We show that both larger apertures and smaller separations between the \ac{LIS} elements render higher resolution images, improving the performance of the system.

\section{Problem formulation}

We consider an industrial scenario where a robot is following a fixed route, and assume that, due to arbitrary reasons, it might deviate from the predefined route and follow an alternative (undesired) trajectory. Hence, our goal is, based on the sensing signal transmitted by the target device, being able to detect whether the robot is following the correct route or not. 

In order to perform the anomalous route detection, we assume that an \ac{LIS} (i.e., a large array of $M$ closely spaced antennas), is placed in the scenario. Therefore, the sensing problem reduces to determine, from the received signal at each of the \ac{LIS} elements, if the transmission has been made from a point at the desired route, denoted by $\mathbf{p}_c\in\mathbb{R}^{3}$, or from an anomalous one, denoted by $\mathbf{p}_a\in\mathbb{R}^{3}$. For the sake of simplicity in a real system implementation, and because we are interested in sensing large scale variations, we resort to the received signal amplitude (equivalently, power). This assumption may lead to simpler system implementations, avoiding the necessity of performing coherent detection. 

A classical approach for the aforementioned problem would be performing a hypothesis test based on the received power signal vector. To that end, consider the received complex signal from either $\mathbf{p}_c$ or $\mathbf{p}_a$ to be
\begin{equation}
    \label{eq:signal}
    \mathbf{y}_k = \mathbf{h}_kx + \mathbf{n}_k, \quad k=\{c,a\},
\end{equation}
with $x$ the  transmitted  (sensing)  symbol, $\mathbf{h}_k\in\mathbb{C}^{M\times 1}$ the channel vector from each point and $\mathbf{n}_k \sim\mathcal{CN}_M(\mathbf{0},\sigma^2\mathbf{I})$ the noise vector. Assume, without loss of generality, that $x=1$. Hence, the received power vector is given by 
\begin{equation}
    \label{eq:PowerDef}
    \mathbf{w}_k = \left(\|y_{1,k}\|^2, \dots, \|y_{M,k}\|^2\right)^T,    
\end{equation}
where $y_{i,k}$ for $i=1,\dots,M$ are the elements of $\mathbf{y}_k$. The hypothesis test is therefore formulated as
\begin{equation}
    \frac{ f_{\mathbf{w}_c}(\mathbf{w} | \mathbf{p}_c) }{ f_{\mathbf{w}_a}(\mathbf{w} | \mathbf{p}_a) } \underset{\mathbf{p}_a}{\overset{\mathbf{p}_c}{\gtrless}} \frac{P_a}{P_c},
\end{equation}
where $f_{\mathbf{w}_k}(\cdot)$ for $k = \{c,a\}$ is the joint probability function of the received signal from each point, $\mathbf{w}$ is the observation vector, and $P_a$ and $P_c$ denote the probability of receiving a signal from $\mathbf{p}_a$ and $\mathbf{p}_c$, respectively. To obtain an optimal estimator, we would need to characterize the joint distribution of the received vector over all the possible anomolaous points, which implies knowing all the possible states of the channels for each path. Also, even in the most simple case, i.e., assuming a pure \ac{LoS} propagation, we would still be unable to distinguish if the two points are in different trajectories or at distinct positions of the same route. Moreover, the a priori probabilities $P_a$ and $P_c$ are needed, which is a non-trivial task. 

In a realistic environment, the complexity of the propagation paths is considerable, and the theoretical analysis becomes cumbersome and site-dependent. Hence, in order to gain insight into how the propagation paths between different positions translate into differences in the received signals, we have to resort on machine learning algorithms. This, together with the use of \ac{LIS}, can provide the necessary information about the propagation environment in order to perform the anomalous route detection.

\section{Holographic sensing}
\label{holImg}

\begin{figure}[t]
\centering
\subfloat[LoS, noiseless. \label{fig:1a}]{\includegraphics[width=0.45\textwidth]{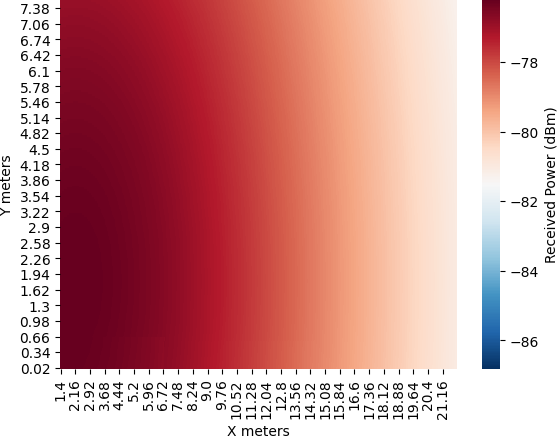}} \hfill
\subfloat[Real scenario, noiseless. \label{fig:1b}]{\includegraphics[width=0.45\textwidth]{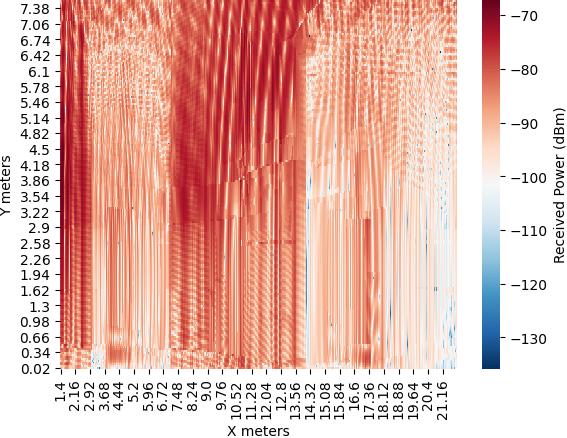}} \hfill
\caption{Holographic images for LOS and Industry scenarios.}
\label{fig:holImages}
\end{figure}
A hologram is a  recorded interference pattern as a result of constructive and destructive combinations of the superimposed light-wavefronts, i.e., a photographic recording of a light field \cite{syms1990practical}. 
In a wireless context, an \ac{LIS} could be described as a structure which uses electromagnetic signals impinging in a determined scatterer in order to obtain a profile of the environment. That is, we can use the signal power received at each of the multiple elements of the \ac{LIS} to obtain a high resolution image of the propagation environment. Using this approach, the complexity of the multipath propagation is reduced to using information represented as an image. This provides a twofold benefit: \emph{i)} the massive number of elements that composed the \ac{LIS} leads to an accurate environment sensing (i.e. high resolution image), and \emph{ii)} it allows the use of computer vision algorithms and image processing techniques to deal with the resulting images. 

As an illustrative example, Fig. \ref{fig:holImages} shows the holographic images obtained from different propagation environments ($x$ and $y$ correspond to the physical dimension of the \ac{LIS}). Specifically, Figs. \ref{fig:1a} correspond to a \ac{LoS} propagation (no scatterers), whilst Fig. \ref{fig:1b} is obtained from an industrial scenario with a rich scattering. Note that, in the case in which different scatterers are placed, their position and shapes are captured by the \ac{LIS} and represented in the image. To the best of the authors' knowledge, this is the first time that imaged-based sensing is proposed in the literature. 

\section{Machine learning for holographic sensing}

\subsection{Model description}

 We here propose the use of a machine learning model to perform the anomalous route classification task, based on the holographic images obtained at the \ac{LIS}. In our considered problem, the training data is obtained by sampling the received power at certain temporal instants while the target device is moving along the route. In order to reduce both training time and scanning periods, we resort on transfer learning \cite{pan2009survey}. Thus, a small dataset can be used, improving the flexibility of the system in real deployments. 
 Among the available strategies for this matter, we will use feature representation.

One of the main requirements for transfer learning is the presence of models that perform well on already defined tasks. These models are usually shared in the form of a large number of parameters/weights the model achieved while being trained to a stable state \cite{sarkar2018hands}. The famous deep learning Python library, Keras \cite{chollet2015keras}, provides an easy way to reuse some of these popular models.
We propose the use of a \ac{SVM} binary classifier, which has been proved to perform correctly when using a large number of features \cite{bishop2006pattern}. In our case, we choose the VGG19 architecture \cite{simonyan2014very}.

\begin{figure}[t!]
    \centering
         \includegraphics[width=\columnwidth]{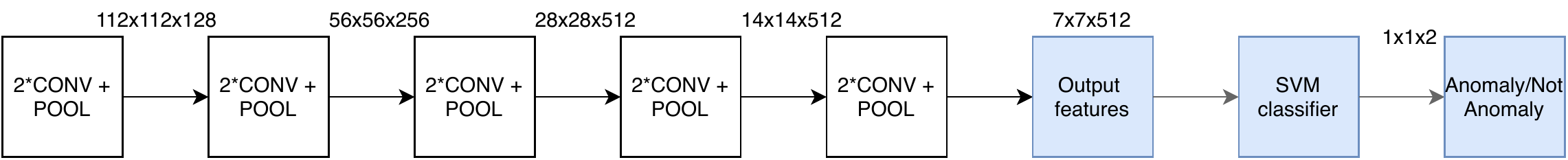}
         \caption{Proposed model. White and blue blocks refer to VGG19 re-used original architecture and to the additional blocks for our task, respectively.}\label{fig:featExt}
\end{figure}

The model is detailed in Fig. \ref{fig:featExt}. In order to perform the feature extraction, we remove the last \ac{FC} that performs the classification for the purpose of VGG19 and modify it for our specific classification task (anomaly/not anomaly in robot's route). We note that the architecture has been frozen for our case, i.e., the weights and biases in VGG19 are fixed and re-used to generate the features to feed the SVM classifier while the regularization parameter $C$ is tuned to prevent overfitting along the training process.

\subsection{Dataset format}
\label{sec:Dataset}
The dataset is obtained by sampling the received signal power at each element of the \ac{LIS} while the robot moves along the trajectories. Formally, we can define the trajectories as the set of points in the space $\mathbf{P}_t\in\mathbb{R}^{N_p\times 3}$ being $N_p$ the total number of points in the route. Let assume the system is able to obtain $N_s$ samples at each channel coherence interval $\forall \;\mathbf{p}_j \in \mathbf{P}_t$, being $\mathbf{p}_j$ for $j=1,\dots,N_p$ an arbitrary point of the route. Hence, the dataset is conformed by $T=N_p\times N_s$ samples (monocromatic holographic image snapshots of received power). 
Each sample is a gray-scale image which is obtained by mapping the received power into the range of [0, 255]. To that end, we apply min-max feature scaling, in which the value of each pixel $m_{i,j}$ for $i=1,\dots,M$ and $j=1,\dots,N_p$ is obtained as 
\begin{equation}
    \label{eq:pixelmapping}
    m_{i,j} = \ceil* {m_{\textsc{min}} + \frac{(w_{i,j} - w_{\textsc{min},j})(m_{ \textsc{max}}-m_{ \textsc{min}})}{w_{ \textsc{max},j} - w_{ \textsc{min},j}}},
\end{equation}
where $w_{i,j}$ are the elements of $\mathbf{w}_j$ in \eqref{eq:PowerDef}, i.e. $w_{i,j}=\|h_{i,j} + n_{i,j}\|^2$, $m_{\textsc{max}} = 255$ and $m_{\textsc{min}} = 0$, and 
\begin{equation}
     w_{ \textsc{max},j}=\max_{\{i=1,...,M\}}\mathbf{w}_{i,j}, \quad  w_{ \textsc{min},j}=\min_{\{i=1,...,M\}}\mathbf{w}_{i,j}
 \end{equation}
  are the maximum and minimum received power value from a point $\mathbf{p}_j$ along the surface.  


The input structure supported by VGG19 is a RGB image of $n_c = 3$ channels. Due to our monocromatic measurements, our original gray-scale input structure is a one-channel image. To solve this problem, we expand the values by copying them into a $n_c = 3$ channels input structure.


Once the feature extraction is performed, the output is $n_c = 512$ channels of size $n_w = 7$ and $n_h = 7$ pixels. Since \ac{SVM} works with vectors, the data is reshaped into an input feature vector formed by $7\times7\times512=25088$ features, meaning our dataset is $\{x^{(i)},y^{(i)}\}_{i=1}^T$, where $x^{(i)}$ is the $i$-{th} $n$-dimensional training input features vector (being $n=25088$), $x^{(i)}_j$ is the value of the $j$-{th} feature, and $y^{(i)}$ is the corresponding desired output label vector.

\section{Model validation}

In order to validate the proposed method, we carried out an extensive set of simulations to analyze the performance of the system. 
To properly obtain the received power values, we use a ray tracing software, therefore capturing the effects of the multipath propagation in a reliable way. Specifically, we consider \textsc{Altair Feko Winprop} \cite{winprop}. 

\begin{figure}[t]
\centering
\subfloat[Use case scenario. \label{fig:scenario1}]{\includegraphics[width=0.48\columnwidth]{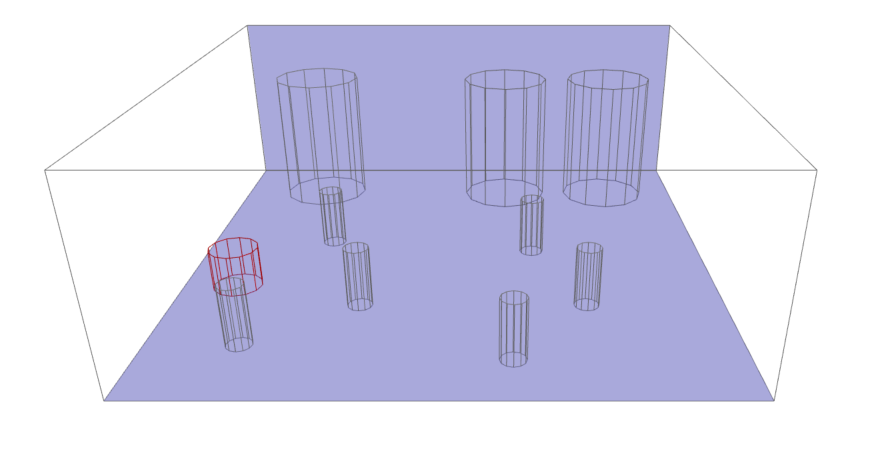}} \hfill
\subfloat[Correct robot route (blue) vs anomalous routes (orange). \label{fig:scenario2}]{\includegraphics[width=0.48\columnwidth]{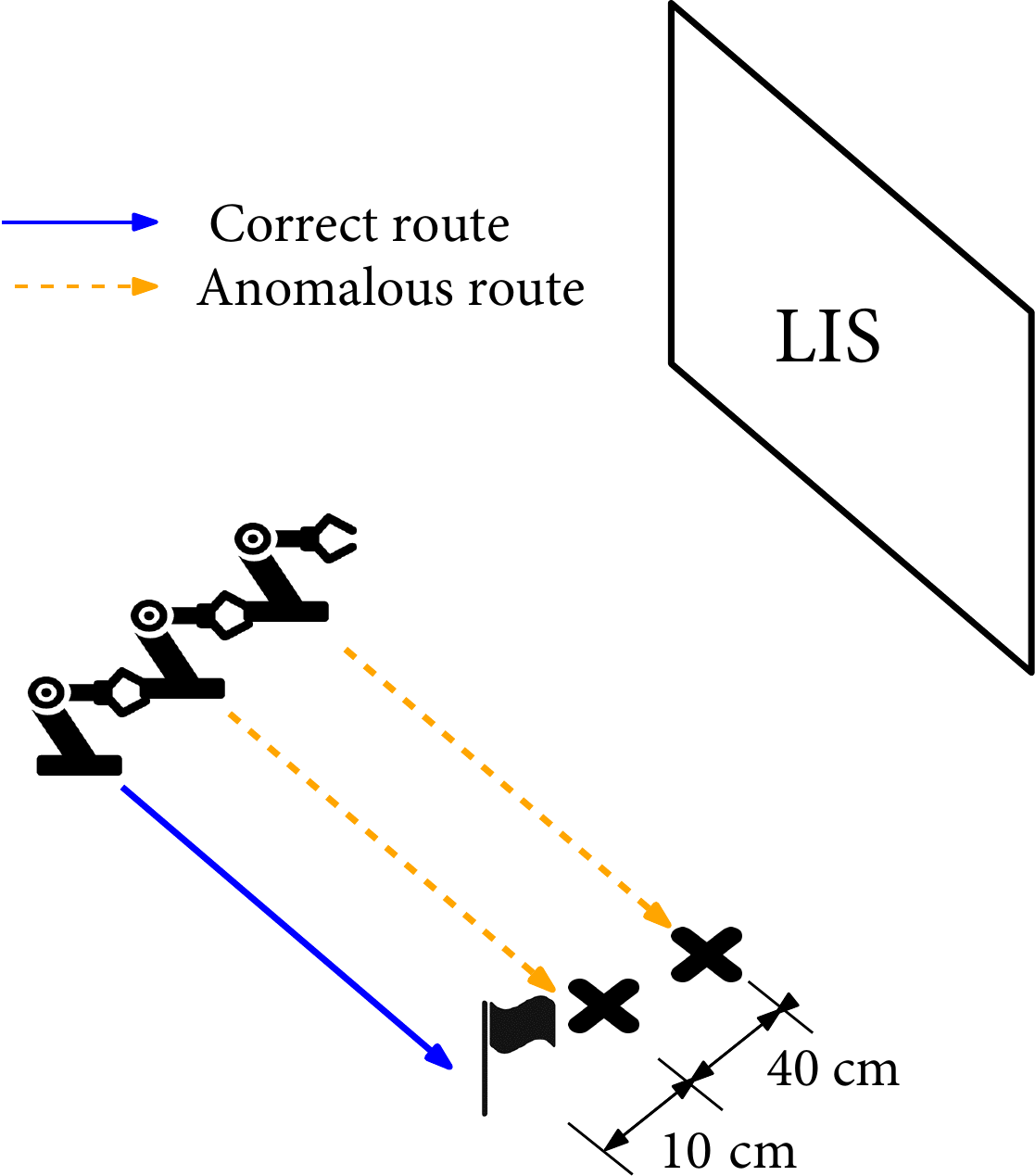}} 
\caption{Simulated scenario.}
\label{fig:scenario}
\end{figure}

\subsection{Simulated scenario}

The baseline set-up is described in Fig. \ref{fig:scenario1}, a small size industrial scenario of size 484 $m^2$. We address the detection of the deviation of the target robot (highlighted in red color) when following a fixed route parallel to the bottom wall, in which the \ac{LIS} is deployed. The distance between the \ac{LIS} and the desired trajectory is $13.9$ m.
For the anomalous routes, a separation of $50/10$ cm have been simulated to analyze the performance of the system when $\Delta d >> \lambda $ and $\Delta d \approx \lambda $ respectively, as detailed in Fig. \ref{fig:scenario2}.

\begin{table}[h!]
\caption{Parameters}
\label{table_parameters}
\centering
\begin{tabular}{|c|c|c|c|c|c|}
\hline
\begin{tabular}[c]{@{}c@{}}Frequency \\ (GHz)\end{tabular} & \begin{tabular}[c]{@{}c@{}}Tx \\ Power \\ (dBm)\end{tabular} & \begin{tabular}[c]{@{}c@{}}Nray \\ paths\end{tabular} & \begin{tabular}[c]{@{}c@{}}Antenna \\ type\end{tabular} & \begin{tabular}[c]{@{}c@{}}Antenna \\ Spacing (cm)\end{tabular} & \begin{tabular}[c]{@{}c@{}}Propagation \\ model \end{tabular} \\ \hline
3.5                                                        & 20                                                           & 20                                                    & Omni                                                                                                                                                                                            & $\frac{\lambda}{2}/\lambda/2\lambda$            & Free Space                                                  \\ \hline
\end{tabular}
\end{table}

For these routes, we simulate in the ray tracing software $N_p$ points, which corresponds to different positions of the robot in both the correct and anomalous routes. Then, $N_s$ holographic image snapshots of the measurements are taken at every $\mathbf{p}_j$, $j = 1, \dots, N_p$. The most relevant parameters used for simulation are summarized in Table \ref{table_parameters}.

In our simulations, we set $N_p=367$ and $N_s=10$, thus the dataset is composed of $T=N_p\times N_s=3670$ radio propagation snapshots containing images of both anomalous and non-anomalous situations, as described in Section \ref{sec:Dataset}. The dataset is split into a 80\% training set and 20\% for the test set. During the training phase, the obtained optimum regularization value is $C=0.001$, which was identified by using a 5-fold cross-validation strategy \cite{anguita2009k}.

\subsection{Received power and noise modeling}

The complex electric field arriving at the $i$-th antenna element at sample time $t$, $\widetilde{E}_{i}(t)$, can be regarded as the superposition of each path, i.e.\footnote{Note that the electric field also depends on the point $\mathbf{p}_j$. However, for the sake of clarity, we drop the subindex $j$ throughout the following subsections.}, 
\begin{equation}
    \label{eq:Esum}
  \widetilde{E}_{i}(t) = \sum_{n=1}^{N_r} \widetilde{E}_{i,n}(t) = \sum_{n=1}^{N_r}E_{i,n}(t) e^{j\phi_{i,n}(t)},  
\end{equation}
where $N_r$ is the number of paths and $\widetilde{E}_{i,n}(t)$ is the complex electric field at $i$-th antenna from $n$-th path, with amplitude $E_{i,n}(t)$ and phase $\phi_{i,n}(t)$. From \eqref{eq:Esum}, and assuming isotropic antennas, the complex signal at the output of the $i$-th element is therefore given by
\begin{equation}
    \label{eq:complexSignal}
    y_i(t) = \sqrt{\frac{\lambda^2Z_i}{4\pi Z_0}} \widetilde{E}_{i}(t) + n_i(t),
\end{equation}
with $\lambda$ the wavelength, $Z_0 = 120\pi$ the free space impedance,  $Z_i$ the antenna impedance, and $n_i(t)$ is complex Gaussian noise with zero mean and variance $\sigma^2$. Note that \eqref{eq:complexSignal} is exactly the same model than \eqref{eq:signal}; the only difference is that we are explicitly denoting the dependence on the sampling instant $t$. For simplicity, we consider $Z_i = 1\,\forall\, i$. Thus, the power $w_i(t) = \|y_i(t)\|^2$ is used at each temporal instant $t$ to generate the holographic image, as pointed out before. Finally, in order to test the system performance under distinct noise conditions, the average \ac{SNR} over the whole route, $\overline{\gamma}$, is defined as\footnote{This is equivalent to average over all the points $\mathbf{p}_j$ of the trajectory $\mathbf{P}$.}
\begin{equation}
    \overline{\gamma} \triangleq  \frac{\lambda^2}{4\pi Z_0 M T\sigma^2}\displaystyle\sum_{t=1}^{T}\sum_{i=1}^{M} |\widetilde{E}_{i}(t)|^2,
\end{equation}
where $M$ denotes the number of antenna elements in the \ac{LIS}.

\subsection{Noise averaging strategy}

Noise is critical in image classification performance \cite{roy2018effects}. Normally, in the image processing literature, noise removal techniques assume additive noise in the images \cite{moeslund2012introduction}, which is not the case in our system. 

Referring to (\ref{eq:signal}) and (\ref{eq:complexSignal}), since we are considering only received powers, the signal at the output of the $i$-th antenna detector is given by 
\begin{equation}
    w_{i} = \left\|\sqrt{\frac{\lambda^2Z_i}{4\pi Z_0}} \widetilde{E}_{i} + n_i\right\|^2,
\end{equation}
where we have dropped the dependence on $t$.  Also, let assume the system is able to obtain $S$ extra samples at each channel coherence interval $\forall\; \mathbf{p}_j \in \mathbf{P}$. That is, at each point $\mathbf{p}_j$, the system is able to get $N'_{s} = N_s \times S$ samples. Since the algorithm only expects $N_s$ samples from each point, we can use the extra samples to reduce the noise variance at each pixel. To that end, the value of each pixel $m_{i,j}$ is not computed using directly $w_{i,j}$ as in \eqref{eq:pixelmapping} but instead
\begin{equation}
    w'_{i,j} = \frac{1}{S}\sum_{s=1}^{S} w_{i, j, s},
\end{equation}
where $ w_{i, j, s}$ denote the received signal power at each extra sample $s=1,\dots,S$. Note that, if $S\to\infty$, then 
\begin{equation}
    \left. w'_{i,j}\right|_{S\rightarrow\infty} = \mathbb{E}[w_{i,j} | h_{i,j}] = \|h_{i,j}\|^2 + \sigma^2,
\end{equation}
meaning that the noise variance at the resulting image has vanished, i.e., the received power at each antenna (conditioned on the channel) is no longer a random variable. Observe that the image preserves the pattern with the only addition of an additive constant factor $\sigma^2$. This effect is only possible if the system would be able to obtain a very large number $S$ of samples within each channel coherence interval. 

\subsection{Performance metrics}

To evaluate the prediction effectiveness of our proposed method, we resort on common performance metrics that are widely used in the related literature. Concretely, we are focusing on the F1-Score which is a metric based on the Precision and  Recall metrics \cite{powers2003recall} and is described as:

\begin{itemize}
\item  Positive F1-Score ($PF_{1}$) and Negative F1-Score($NF_{1}$) as the harmonic mean of precision and recall:
\begin{align}
\text{$PF_{1}$} &=  2\cdot\frac{\text{PP}\cdot \text{RP}}{\text{PP + RP}}, & \text{$NF_{1}$} &= 2\cdot\frac{\text{PN}\cdot \text{RN}}{\text{PN + RN}}. 
\end{align}
\end{itemize}
Where $\text{PP}$ and $\text{RP}$ stand for Precision and Recall of the positive class (anomaly) while $\text{PN}$ and $\text{RN}$ stand for Precision and Recall of the negative class (not anomalous situation).

\section{Numerical results and Discussion}


Generally, in the considered industrial setup, it would be more desirable to avoid undetected anomalies (which may indicate some error in the robot or some external issue in the predefined trajectory) than obtaining a false positive. Hence, all the figures in this section shows the algorithm performance in terms of the $PF_{1}$ metric. 

\subsection{Impact of sampling and noise averaging}

To evaluate the impact of both sampling and noise averaging, we consider an \ac{LIS} compounded by $M=128\times128$ antennas and a spacing $\Delta s=\lambda/2$ for the $\Delta d=50$ cm anomalous route. 

\begin{figure}[h!]
    \centering
    \includegraphics[width=0.7\columnwidth]{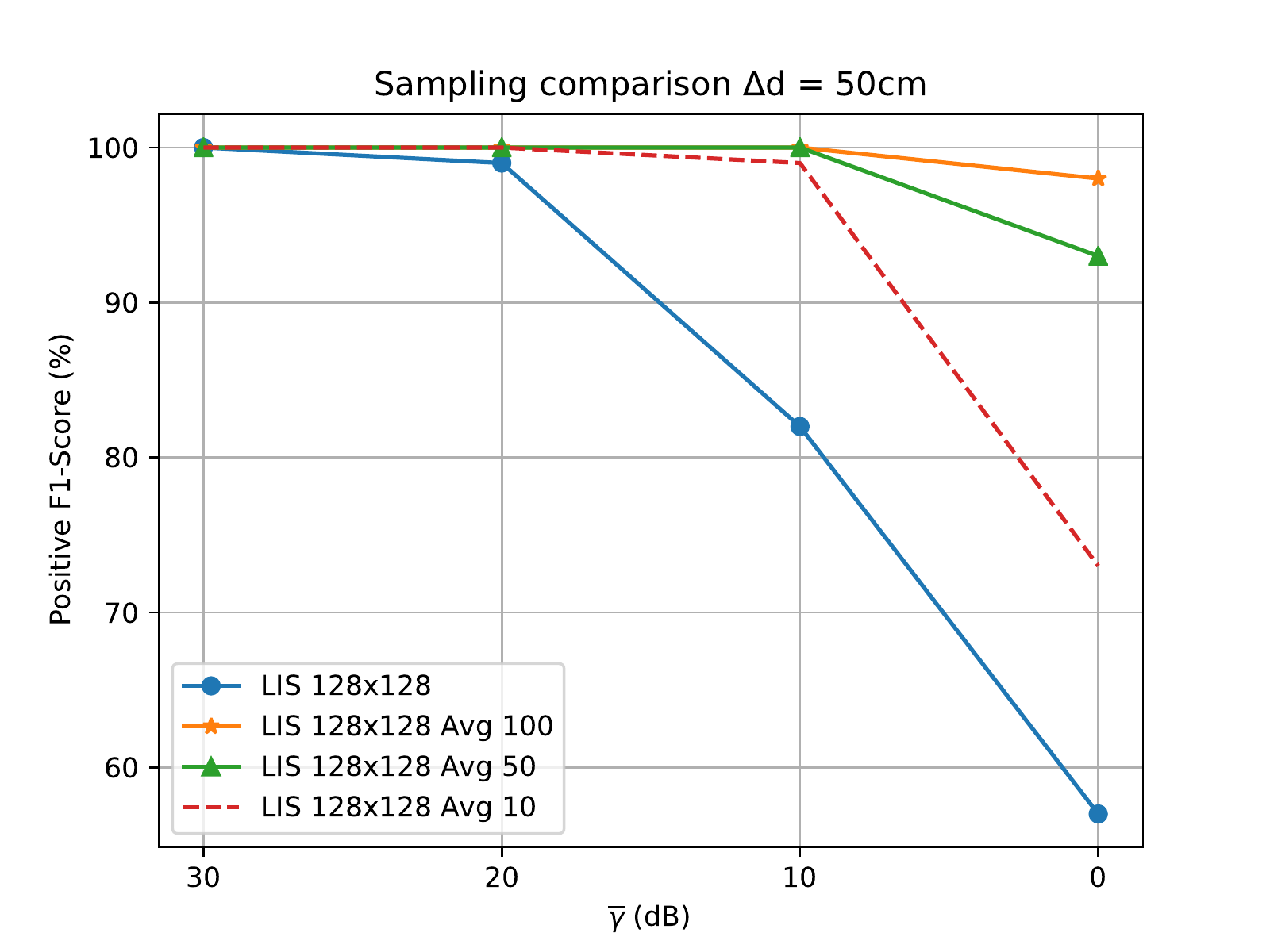}
    \caption{$PF_1$ score averaged noise vs non-averaged.}\label{fig:avg_vs_noavg}
\end{figure}

For our particular case, $N_s'\in \{1000, 500, 100\}$. Then $\forall \; \mathbf{p}_j$ we use $S=\frac{N_s'}{N_s}$ samples for obtaining $N_s$ $S$-averaged samples for training the algorithm, being still $T=N_p\times N_s=3670$. Note that the number of samples $N_s'$ would depend on the sampling frequency and the second order characterization of the channel, i.e., the channel coherence time and its autocorrelation function. 

Figure \ref{fig:avg_vs_noavg} shows the performance of the system when using non-averaged samples and averaged ones respectively. The blue line represents the system when non-averaged data is being used. When the noise contribution is non-negligible in the interval $\overline{\gamma}\in[10\text{ dB}, 0\text{ dB}]$, the detection performance presents a significant drop. Thanks to the averaging, results are significantly improved, even in the critical interval. As expected, when noise level is higher, more samples are needed to preserve the pattern by averaging, being $N_s'=1000$ the one which yields a better performance. For the following discussions, this sampling strategy will be used, meaning we are using $S=100$ extra samples.

\subsection{Impact of antenna spacing}

To evaluate the impact of inter-antenna distance, we fix the aperture to $5.44\times5.44$ m, we assess the performance in both $\Delta d=50/10$ cm, and we analyze different spacings with respect to the wavelength ($\lambda/2$, $\lambda$ and $2\lambda$). 

\begin{figure}[h!]
    \centering
    \includegraphics[width=0.7\columnwidth]{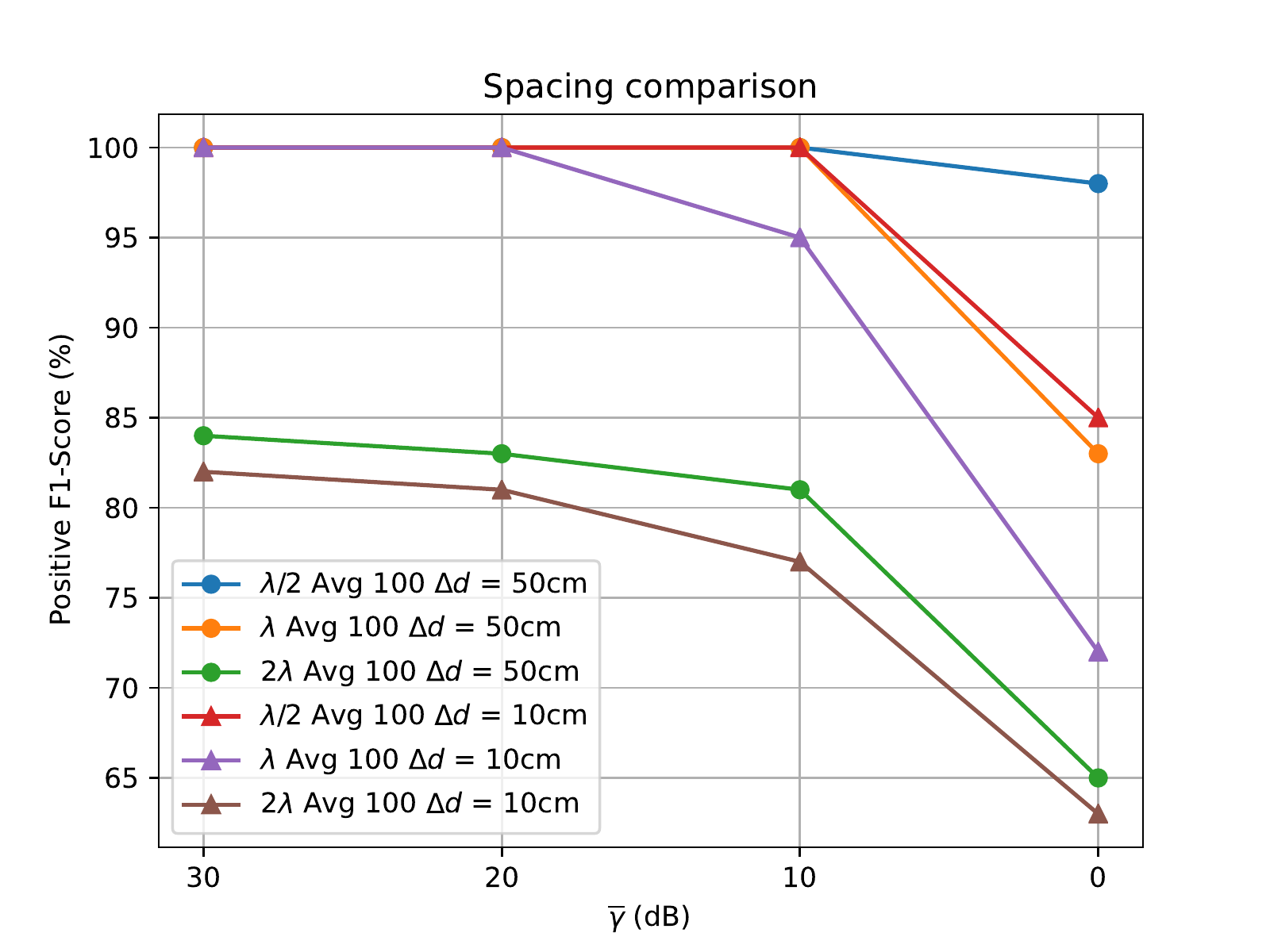}
    \caption{$PF_1$ score antenna spacing}\label{fig:spacings}
\end{figure}

The performance results for the distinct configurations are depicted in Fig. \ref{fig:spacings}. As observed, the spacing of $2\lambda$ --- which is far from the concept of \ac{LIS} --- is presenting really inaccurate results showing that the spatial resolution is not enough. We can conclude that the quick variations along the surface provide important information to the classifier performance. Besides, this information becomes more important the lower the distance between the routes is. The performance drop due to the closer distances among the routes is related to the pattern classification. The closer the routes are, the more similar the pattern is making more challenging to perform the detection. However, reducing the antenna spacing even more can improve the performance when routes are even closer. What is more, the effect of antenna densification for a given aperture is highlighted and it can be seen that the lowest spacing leads to the best results.

\subsection{LIS aperture comparisons}

In this case \ac{LIS} with different apertures have been evaluated. The spacing is fixed to $\lambda/2$.

 \begin{figure}[h!]
    \centering
    \includegraphics[width=0.7\columnwidth]{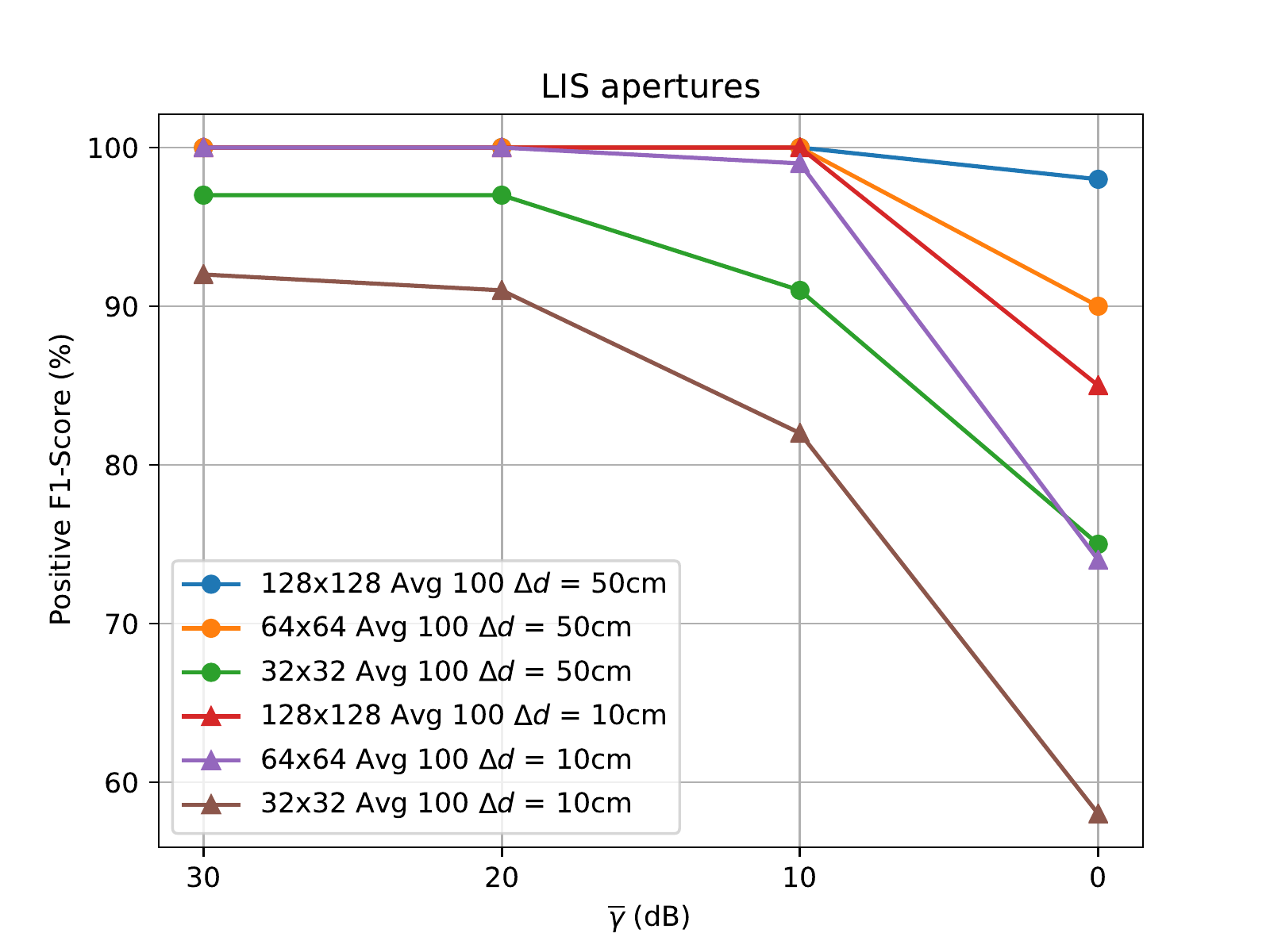}
    \caption{Different LISs apertures comparison}\label{fig:lambda/2aperturecomparison}
\end{figure}

Looking at Fig. \ref{fig:lambda/2aperturecomparison}, the aperture plays a vital role in the sensing performance. Increasing the number of antennas leads to a higher resolution image, being able to capture the large-scale events occurring in the environment more accurately. Note the usage of incoherent detectors is yielding to a good performance when the aperture is large enough. The key feature for this phenomena is the \ac{LIS} pattern spatial consistency, i.e., the ability of representing the environment as a continuous measurement image.

\section{Conclusions}
We have shown the potential of LIS for sensing the environment, being able to provide high resolution radio images of the propagation environment that can be processed by existing and versatile solutions in the context of computer vision algorithms. This sensing technique, which we consider appropriate to refer to as holographic sensing, arises as a robust solution to capture the large scale events of a target scenario, with the inherent advantage that the received signal phase does not need to be estimated. The combined usage of both LIS and machine learning algorithms may be potentially used in the context of cognitive radio and multiuser massive MIMO as a support technology to enhance the performance of these systems. 
%
%
%
%

\bibliographystyle{unsrt}
\bibliography{./biblio}   

\end{document}